\documentclass[prb,reprint]{revtex4-2}
\usepackage{graphicx,amssymb,amsfonts,amsmath}
\usepackage{color}
\usepackage{mathtools}
\usepackage{latexsym}
\usepackage{graphicx}
\usepackage{float}
\usepackage{dcolumn}
\usepackage{bm}
\usepackage{amssymb}
\usepackage{amsmath}
\usepackage{mathtools}
\usepackage[space]{grffile}
\usepackage{xcolor}
\usepackage{hyperref}
\hypersetup{
    colorlinks=true,
    linkcolor=blue,
    filecolor=blue,      
    urlcolor=blue,
    citecolor=blue
    }

\begin{document}

\title{Floquet Product Mode}
\author{Hsiu-Chung Yeh$^{1}$}
\author{Achim Rosch$^{2}$}
\author{Aditi Mitra$^{1}$}
\affiliation{
$^{1}$Center for Quantum Phenomena, Department of Physics,
New York University, 726 Broadway, New York, New York, 10003, USA\\
$^{2}$Institute for Theoretical Physics, University of Cologne, 50937 Cologne, Germany
}

\begin{abstract}
Results are presented for the dynamics of edge modes in interacting Floquet Ising chains. It is shown that in addition to the quasi-stable $0$ and $\pi$ edge modes, a third long lived edge mode arising from the operator product of the  $0$ and $\pi$ edge modes exists. Depending on the microscopic parameters, this Floquet product mode is shown to have a substantially longer lifetime than the individual $0$ and $\pi$ modes. 
This is triggered by a scattering process
which converts a $0$ mode into a $\pi$  mode while scattering two bulk excitations. This process can lead to a rapid decay of both $0$ and $\pi$ mode without affecting the product mode.
\end{abstract}
\maketitle
\section{Introduction}
Floquet Ising chains with open boundary conditions, and weak integrability breaking,  host operators localized at the edges known as almost strong modes \cite{Fendley17,Nayak17,Yates19,Yates20a,Parker2019,Yao20,Yates21,Yates22,yeh2023decay,yeh2023Impurity,schmid2024robust,mi2022noise}. These operators have the property that they are quasi-conserved, i.e, their infinite temperature autocorrelations functions are long lived in the theromodynamic limit, with the  lifetime approaching infinity as the size of the  integrability  breaking perturbations are reduced. In addition, one can have two flavors of almost strong modes \cite{Kitaev01,Fendley2012,FendleyXYZ,vernier2024strong,Alicea16,Garrahan18,Garrahan19,Bardarson22,Fendley23,Zoller11,Sen13,Else16,Roy16,Khemani16, Potter16}, those that almost commute with the Floquet unitary, also known as almost  strong zero modes,  and those that almost anti-commute with the Floquet unitary, also known as almost strong $\pi$ modes \cite{Yates19,Yates21,Yates22,yeh2023decay}. 

While dynamics of $0$ and $\pi$ modes are well studied, here we report on a very general observation. When two or more conserved quantities exist, additional conserved quantities can be constructed from the operator product of the individual conserved quantities. For the Floquet Ising model, in the phase where both $0$ and $\pi$ modes exist, we construct the Floquet product mode from the operator product of the strong zero and $\pi$ modes. We show that this is not a trivial object as it can have dynamics which is qualitatively different from the dynamics of the constituent objects. Surprisingly, it is possible for the Floquet product mode to be more stable than the $0$ and $\pi$ modes.

We also show that even when the decay of  the $0,\pi$ edge modes is given by the same functional power of the integrability breaking term, the existence of one mode can strongly modify, the scattering matrix elements, and therefore the magnitude of the decay rate of the other mode. 

The paper is organized as follows. In Section \ref{Sec: II}, the model is presented and its properties in the absence of integrability breaking perturbations are described. In particular, the topological phase diagram is summarized, the expressions for the $0$,$\pi$ strong modes are given, and the product mode is defined. In Section \ref{Sec: III} theoretical predictions for the decay rates of the modes are presented. The regime of parameter space where Fermi Golden Rule (FGR) is valid is identified, with expressions for the FGR decay rates presented. In addition, the leading power of the decay rates in the regime where FGR does not hold, are given. In Section \ref{Sec: IV}, numerical results are presented and compared with the analytic predictions of Section \ref{Sec: III}. Finally, we conclude in Section \ref{Sec: V}, and give more technical details in three appendices. 

\begin{figure}[h!]
    \centering
    \includegraphics[width=0.25\textwidth]{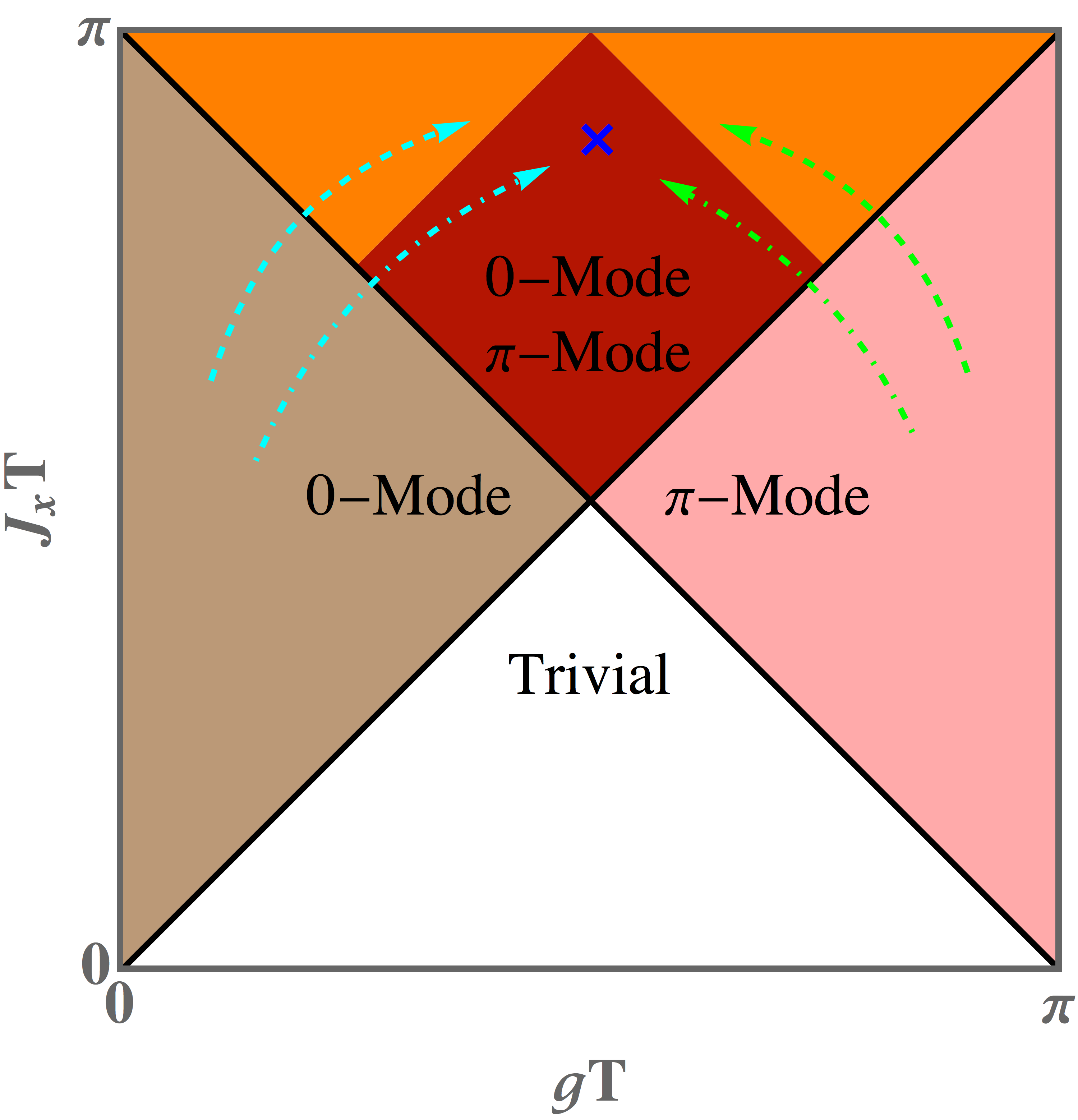}
    \caption{The phase diagram of the unperturbed Floquet system.   In the presence of perturbations, both $0$-mode ($\psi_0$) and $\pi$-mode ($\psi_\pi$) have finite lifetimes in the thermodynamic limit. When both $0$ and $\pi$ modes exist, they can scatter with bulk modes, and also with each other. 
    The two different colors in the $0$-$\pi$ phase represent two different decay rates for $0$-$\pi$ scattering. In addition, when both modes are present, a product mode $\Psi = i\psi_0\psi_\pi$ also exists, whose lifetime, depending on the location in the phase diagram, can be much longer than that of the $0$ and $\pi$-modes (example, the blue cross).  The four dashed curves are chosen such that the localization length of the $0$-mode (left curves, cyan) or the $\pi$-mode (right curves, green) is fixed, but all three modes,   $0,\pi$ and the product mode, show a strong variation in their decay rates along the curves.}
    \label{Fig: Phase Diagram}
\end{figure}

\section{Model}
\label{Sec: II}
We study stroboscopic time-evolution of an open chain of length $L$ according to the Floquet unitary 
\begin{align}
    U = e^{-i\frac{T}{2}J_z H_{zz}}e^{-i\frac{T}{2}g H_z}e^{-i\frac{T}{2}J_x H_{xx}},
    \label{Eq: Full Unitary}
\end{align}
where 
\begin{align}
    H_{xx} = \sum_{i=1}^{L-1} \sigma_i^x\sigma_{i+1}^x;\ H_{z} = \sum_{i=1}^{L} \sigma_i^z;\ H_{zz} = \sum_{i=1}^{L-1} \sigma_i^z\sigma_{i+1}^z.
    \label{Eq: Ising Hamiltonian}
\end{align}
Above $\sigma^{x,y,z}_i$ are Pauli matrices on site $i$,  $g$ is the strength of the transverse-field and $J_{x,z}$ is the strength of the Ising interaction in the $x,z$-direction. $T$ denotes the period where as $T\rightarrow 0$, one recovers Hamiltonian dynamics.   The model has a $Z_2$ symmetry $\mathcal{D} = \sigma^z_1 \ldots \sigma^z_L$. 
For $J_z = 0$, the Floquet unitary $U_0 = U|_{J_z = 0}$ becomes non-interacting, i.e, it can be expressed entirely in terms of Majorana fermion bilinears \cite{Sen13,Yates19,Yates21,yeh2023decay}. In addition,  two types of Majorana edge modes, $0$ and $\pi$-modes ($\psi_0$ and $\psi_\pi$), are allowed, with the phase diagram \cite{Khemani16,Sondhi16a} shown in Fig.~\ref{Fig: Phase Diagram}. These edge modes anti-commute with the $Z_2$ symmetry, and in the thermodynamic limit obey $[U_0,\psi_0] = 0$ and $\{U_0,\psi_\pi\} = 0$, and hence have an infinite lifetime. Their analytic expressions are \cite{yeh2023decay} 
\begin{align}
    &\psi_0 = \mathcal{N}_0 \sum_{l = 1}^L \alpha_l \xi_0^{l-1};
    &\psi_\pi = \mathcal{N}_\pi \sum_{l = 1}^L \beta_l \xi_\pi^{l-1}, \label{psi0pi}
\end{align}
where $\mathcal{N}_0$ and $\mathcal{N}_\pi$ are normalization prefactors. $\alpha_l = \cos(gT/2)a_l + \sin(gT/2)b_l$ and $\beta_l = \sin(gT/2)a_l - \cos(gT/2)b_l$, which are linear combination of Majoranas on the odd ($a_l$) and even ($b_l$) sites, see Appendix \ref{Sec:A}. The localization length of the edge modes are given by $\xi_0 = \tan(gT/2)\cot(J_xT/2)$ and $\xi_\pi = -\cot(gT/2)\cot(J_xT/2)$.  The phase boundaries in Fig.~\ref{Fig: Phase Diagram} correspond to $|\xi_{0/\pi}|=1$, when the modes cannot be normalized. 

When both $0$ and $\pi$ Majorana edge modes are present, there is another mode with an infinite lifetime. We dub this mode  the Floquet product mode because it  is  a product of the two Majorana edge modes, $\Psi=i\psi_0\psi_\pi$. However, the product mode is not a Majorana mode since it does not obey Majorana anticommutation relations. As the perturbation is turned on, $J_z \neq 0$, all these three modes now have finite lifetime. However, these are still long-lived quasi-stable modes, where the $0,\pi$ modes are known as almost strong modes. To probe this phenomena, we use that $0$ and $\pi$-modes are localized on the edge with $\mathcal{O}(1)$ overlap with $\sigma_1^x=a_1$ according to \eqref{psi0pi} and the Jordan Wigner transformation (see details in Appendix \ref{Sec:A}).  As for the product mode, since $i\alpha_1\beta_1 = -ia_1b_1 = -i\sigma_1^x\sigma_1^y =\sigma_1^z$, the product mode is also localized on the edge, but with $\mathcal{O}(1)$ overlap with $\sigma_1^z$.  Thus, these modes can be detected by the following infinite temperature autocorrelation function of $\sigma_1^x$
 \cite{Fendley17,Yates19,Yates20,Yao20,Yates20a,yeh2023decay,yeh2023Impurity}  and $\sigma_1^z$ 
\begin{align}
    A_\infty^x (n) = \frac{1}{2^L} \text{Tr}[\sigma_1^x(n) \sigma_1^x];
    A_\infty^z (n) = \frac{1}{2^L} \text{Tr}[\sigma_1^z(n) \sigma_1^z],
    \label{Eq: AutoCorrelation}
\end{align}
where $n$ is the stroboscopic time-period.

\begin{figure*}
    \centering
    \includegraphics[width=0.23\textwidth]{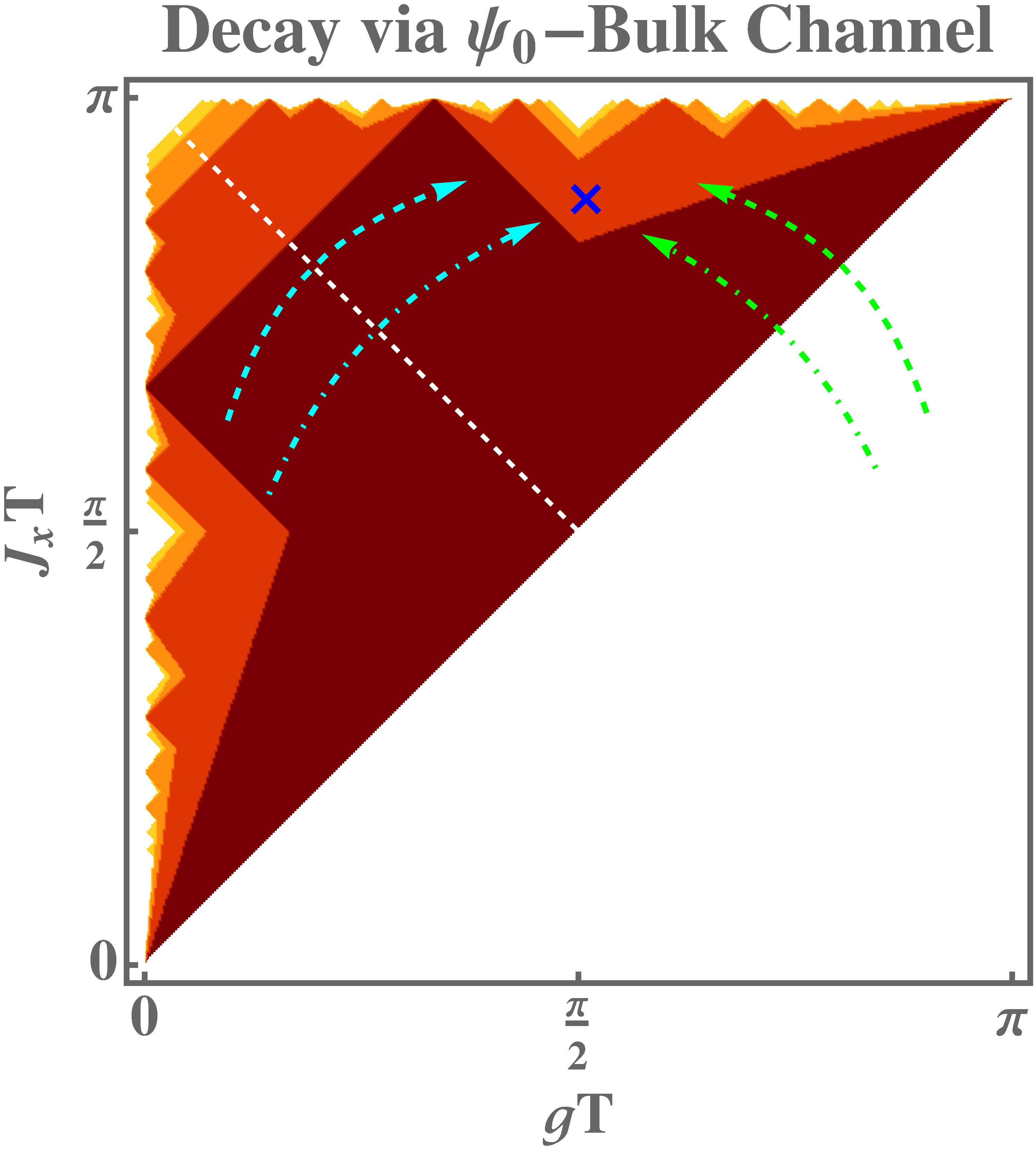}
    \includegraphics[width=0.23\textwidth]{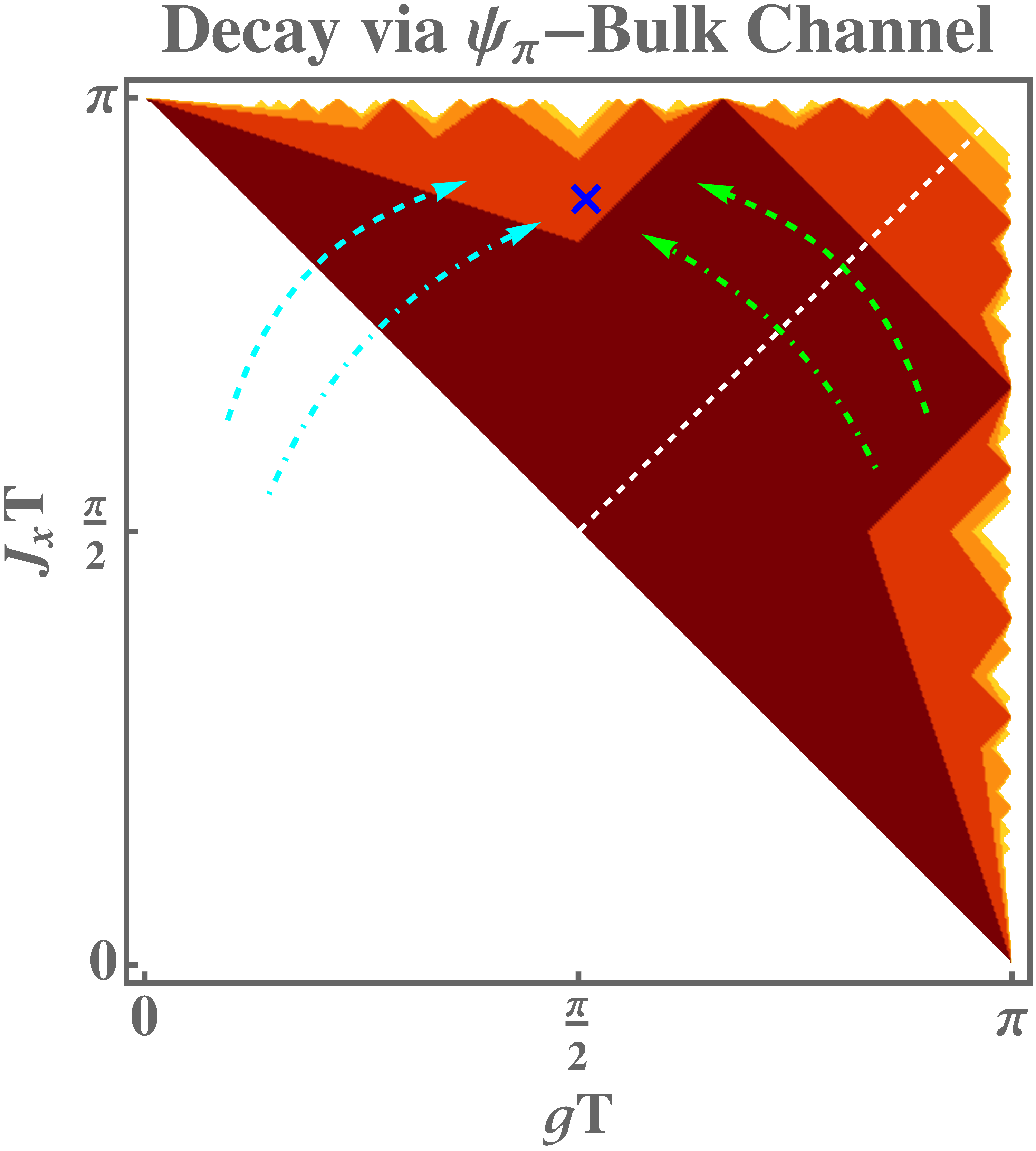}
    \includegraphics[width=0.295\textwidth]{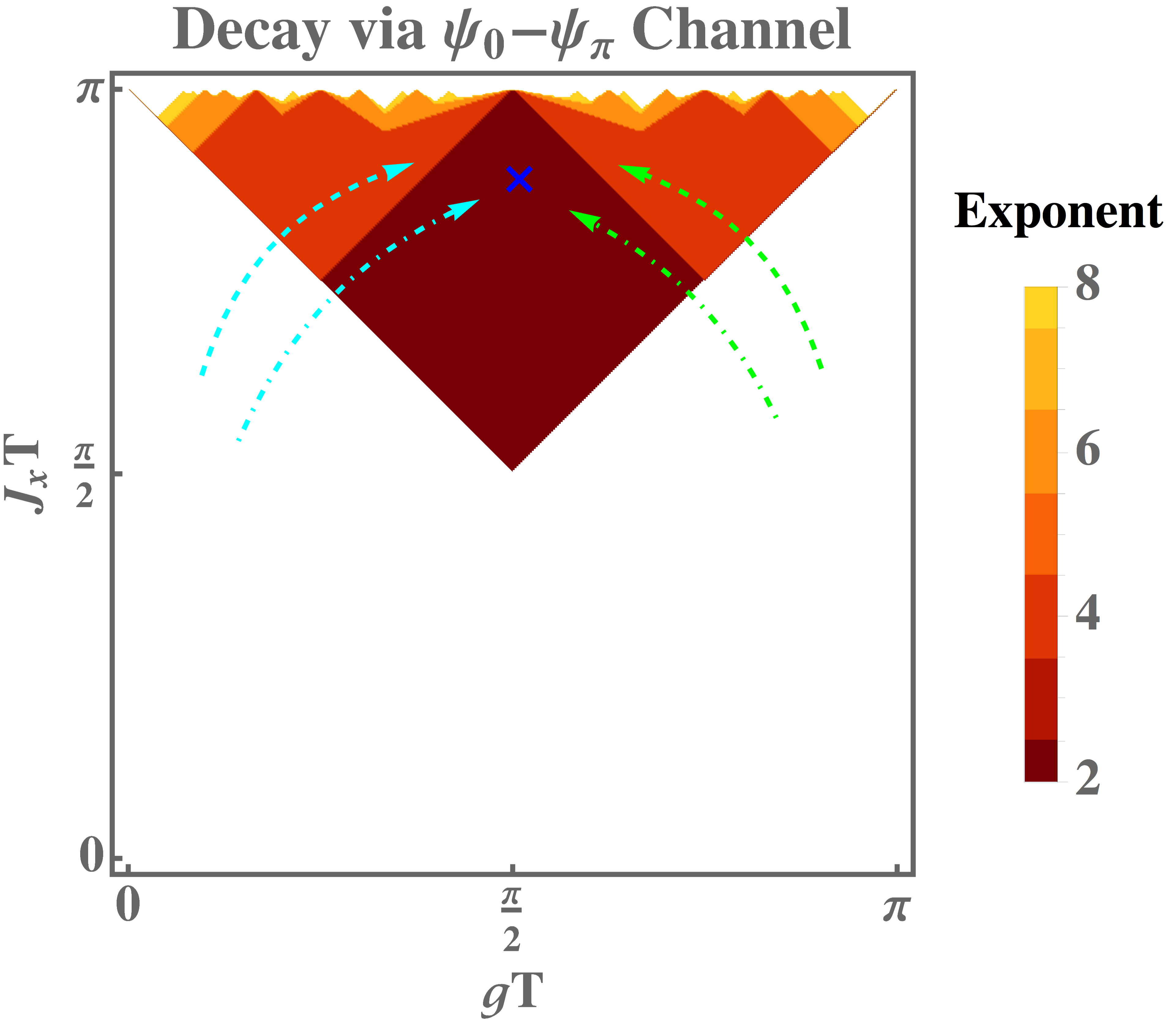}
    \caption{Plot of the exponent $m$ characterizing the decay via three different channels: $0$-mode scatters with bulk excitations (left panel), $\pi$-mode scatters with bulk excitations (middle panel) and the $0$-mode and $\pi$-mode scatter with each other via bulk excitations (right panel). The decay rate is predicted to be proportional to $J_z^m$ for $J_z \to 0$ where the exponent $m$ is encoded in the color of the plot. The exponent $m=2 n$ is determined from the smallest $n$ satisfying the quasi-energy conservation \eqref{Eq: Quasi-energy conservation} for each panel separately. The blue cross highlights the case where the dominant scattering channel is the scattering between the $0$ and $\pi$-modes, and is investigated in Fig.~\ref{Fig: Autocorrelation}. The four dashed curves (identical to Fig.~\ref{Fig: Phase Diagram}) correspond to the case where the scattering between the edge mode and bulk excitations is 2nd order while the scattering between the edge modes ($\psi_0$-$\psi_\pi$ channel) is 2nd order for the inner curves and 4th order for the outer curves. The corresponding FGR calculations are presented in Fig.~\ref{Fig: FGR 0-mode fixed} (left dashed curves) and in Fig.~\ref{Fig: FGR pi-mode fixed} of Appendix \ref{Sec:B} (right dashed curves).}
    \label{Fig: Decay Channel}
\end{figure*}

\section{Fermi Golden Rule and Beyond}
\label{Sec: III}
Denoting the perturbation as
\begin{align}
    V = -\frac{J_z}{2}H_{zz} = -\frac{J_z}{2}\sum_l^L \alpha_l \beta_l \alpha_{l+1} \beta_{l+1},\label{Eq: perturbation}
\end{align}
the lifetime of edge modes can be calculated from perturbation theory in $V$.
The FGR decay rate  $\Gamma_{0(\pi)}$ of the $0(\pi)$ mode is second order in the perturbation, and given by \cite{yeh2023decay} 
\begin{align}
    \Gamma_0 &= \frac{T}{2^L} 
    \biggl( \frac{1}{2}  \text{Tr} \left[ \dot{\psi_0} \dot{\psi_0} \right] + \sum_{n= 1}^{\infty} \text{Tr}\left[\dot{\psi_0}(n) \dot{\psi_0}\right] \biggr)\nonumber\\
    &= \frac{1}{2^L} \sum_{i,j} |\langle i| \Dot{\psi_0} |j\rangle|^2  \pi \delta_F\left(\epsilon_i -\epsilon_j \right);\\
    \Gamma_\pi &= \frac{T}{2^L} 
    \biggl( \frac{1}{2}  \text{Tr} \left[ \dot{\psi_\pi} \dot{\psi_\pi} \right] + \sum_{n= 1}^{\infty} (-1)^n\text{Tr}\left[\dot{\psi_\pi}(n) \dot{\psi_\pi}\right] \biggr)\nonumber\\ &= \frac{1}{2^L} \sum_{i,j} |\langle i| \Dot{\psi_\pi} |j\rangle|^2  \pi \delta_F\left(\epsilon_i -\epsilon_j + \frac{\pi}{T} \right),
\end{align}
where $\Dot{\psi}_{0/\pi} = i[V,\psi_{0/\pi}]$. We define $\Dot{\psi}_{0/\pi} (n) = (U_0^\dagger)^n \Dot{\psi}_{0/\pi} U_0^n$, $| i \rangle$ are the many-particle eigenstates of the unperturbed Floquet unitary $U_0$ with eigenvalues $e^{-i \epsilon_i T}$, and the $\delta_F$ function encodes energy conservation modulo $2 \pi/{T}$, with $\delta_F(\epsilon)=\sum_m \delta(\epsilon+m 2\pi/T)$.
The matrix element is determined by the norm square of the commutator between the edge mode and the perturbation $V$. The Majoranas $\{a_l\}$ and $\{b_l\}$ are superpositions of Majorana edge modes and bulk degrees of freedom: $\alpha_l = (\psi_0|\alpha_l)\psi_0 + \Tilde{\alpha}_l$ and $\beta_l = (\psi_\pi|\beta_l)\psi_\pi + \Tilde{\beta}_l$, where $\Tilde{\alpha}_l$ and $\Tilde{\beta}_l$ denote  bulk degrees of freedom. The inner product between two operators is defined as $(A|B) = \text{Tr}[A^\dagger  B]/2^L$. For a non-zero commutation of the $0$-mode ($\pi$-mode), $[V,\psi_0] \neq 0$ ($[V,\psi_\pi] \neq 0$), one requires one of the four Majoranas in \eqref{Eq: perturbation} to be a $0$-mode ($\pi$-mode) and the others could be a $\pi$-mode (0-mode) or bulk degrees of freedom. Therefore, one can further separate the FGR decay into different channels. For example, the perturbing term  $\psi_0\Tilde{\beta}\Tilde{\alpha}\Tilde{\beta}$ corresponds to $0$-mode scattering with three bulk modes, while the term $\psi_0\psi_\pi\Tilde{\alpha}\Tilde{\beta}$  corresponds to the $0$-mode scattering with the $\pi$-mode and two bulk modes. Thus the FGR decay rate of the $0,\pi$ modes can be split into the following scattering channels: $\psi_0$-bulk, $\psi_\pi$-bulk and $\psi_0$-$\psi_\pi$ channel.  
However, the FGR decay of the product mode only involves the  $\psi_0$-bulk and $\psi_\pi$-bulk channel since $[\Psi,\psi_0\psi_\pi\Tilde{\alpha}\Tilde{\beta}]=0$. Thus, in 2nd order, the product mode decays when the $0$ and $\pi$ modes decay independently by bulk channels. 

\begin{figure*}
    \centering
    \includegraphics[width=0.32\textwidth]{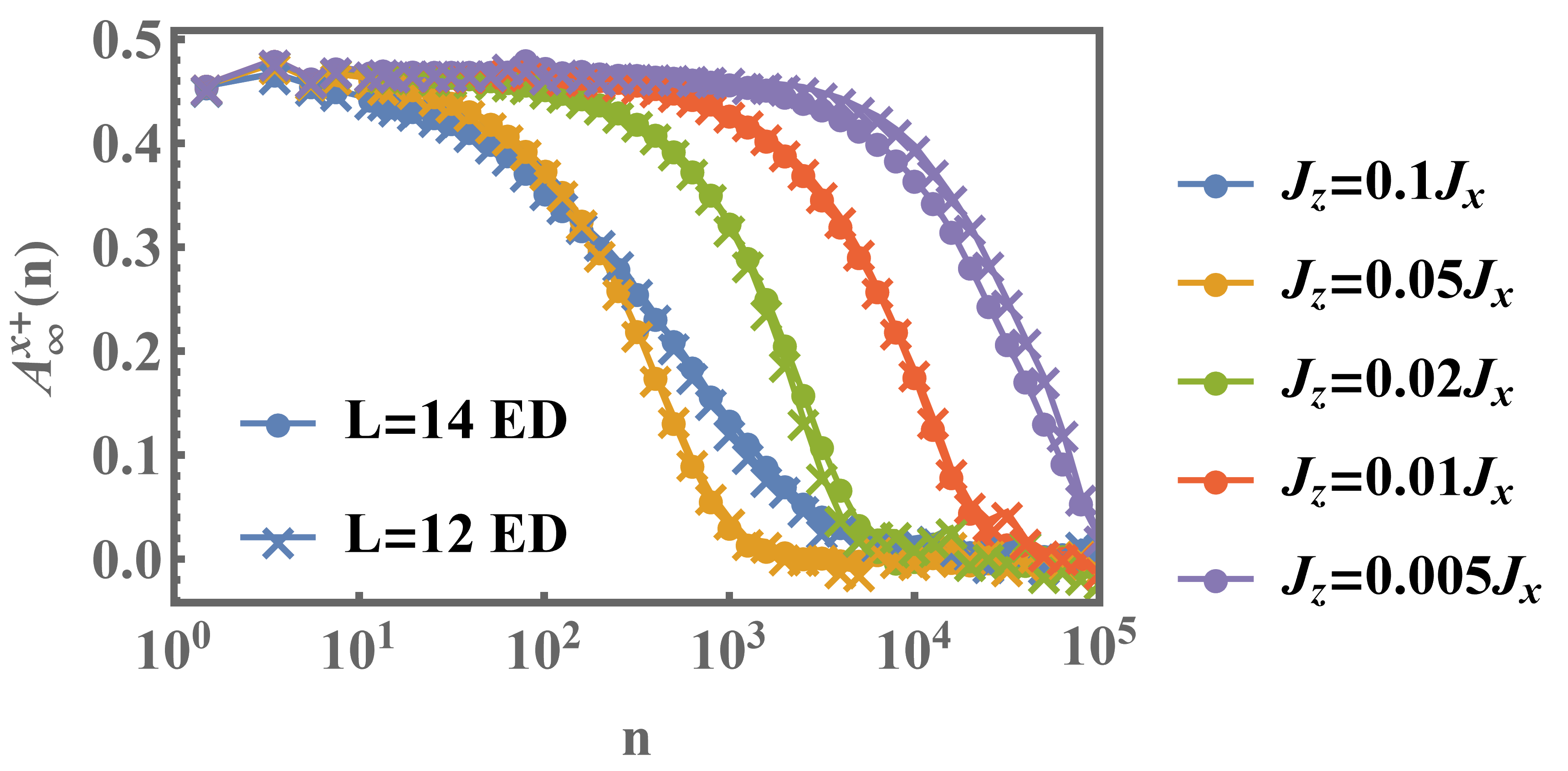}
    \includegraphics[width=0.32\textwidth]{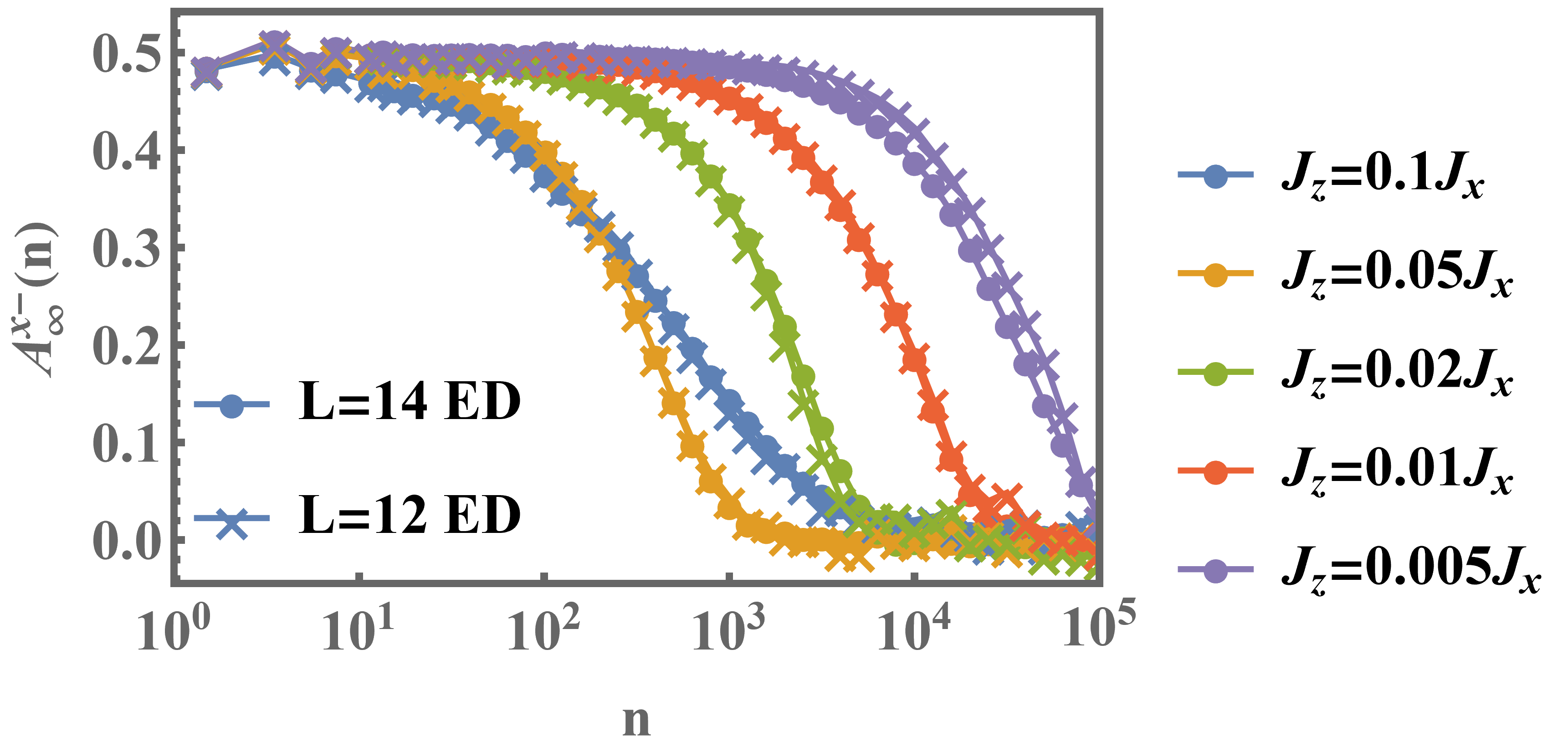}
    \includegraphics[width=0.32\textwidth]{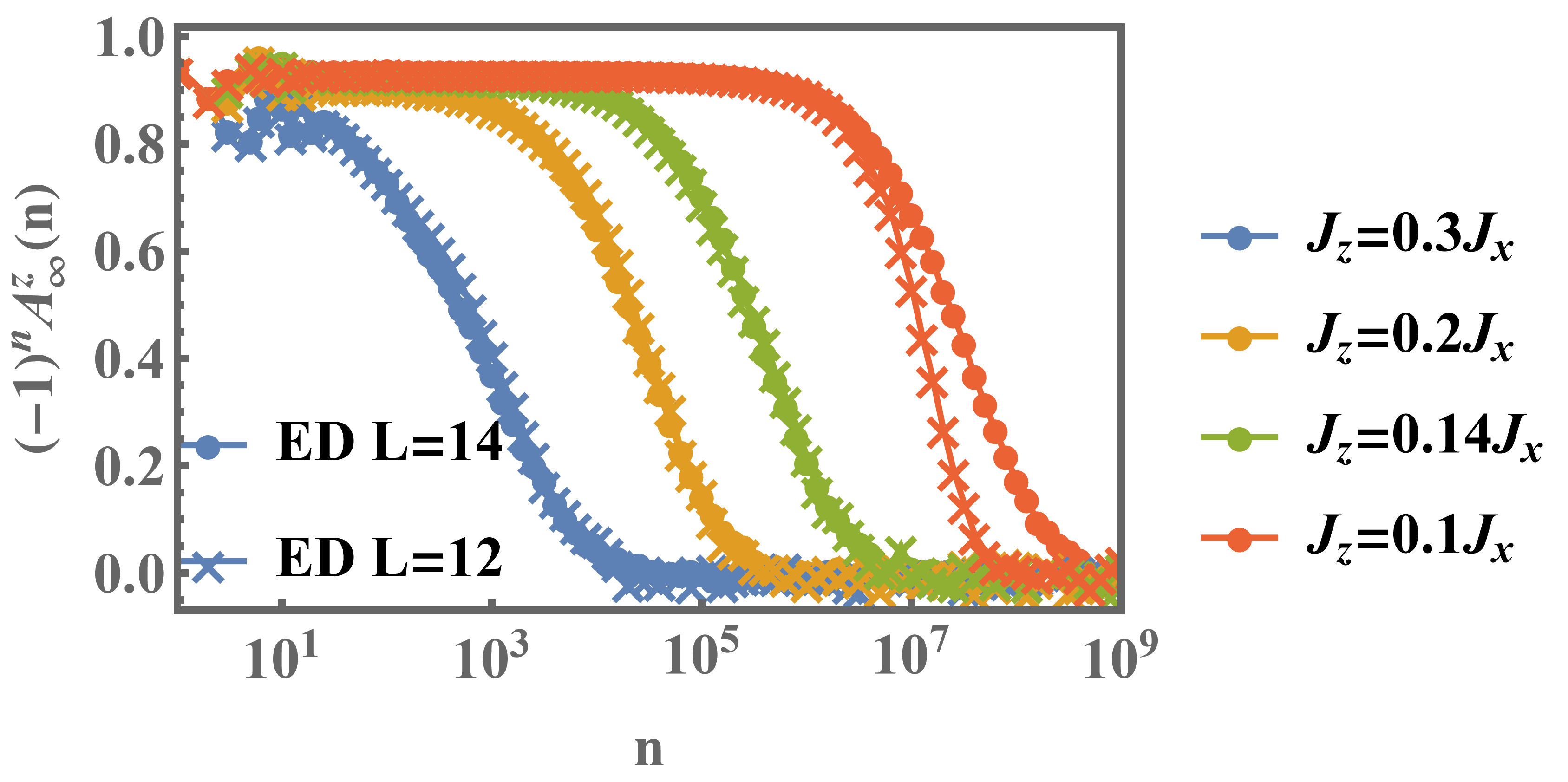}
    \includegraphics[width=0.32\textwidth]{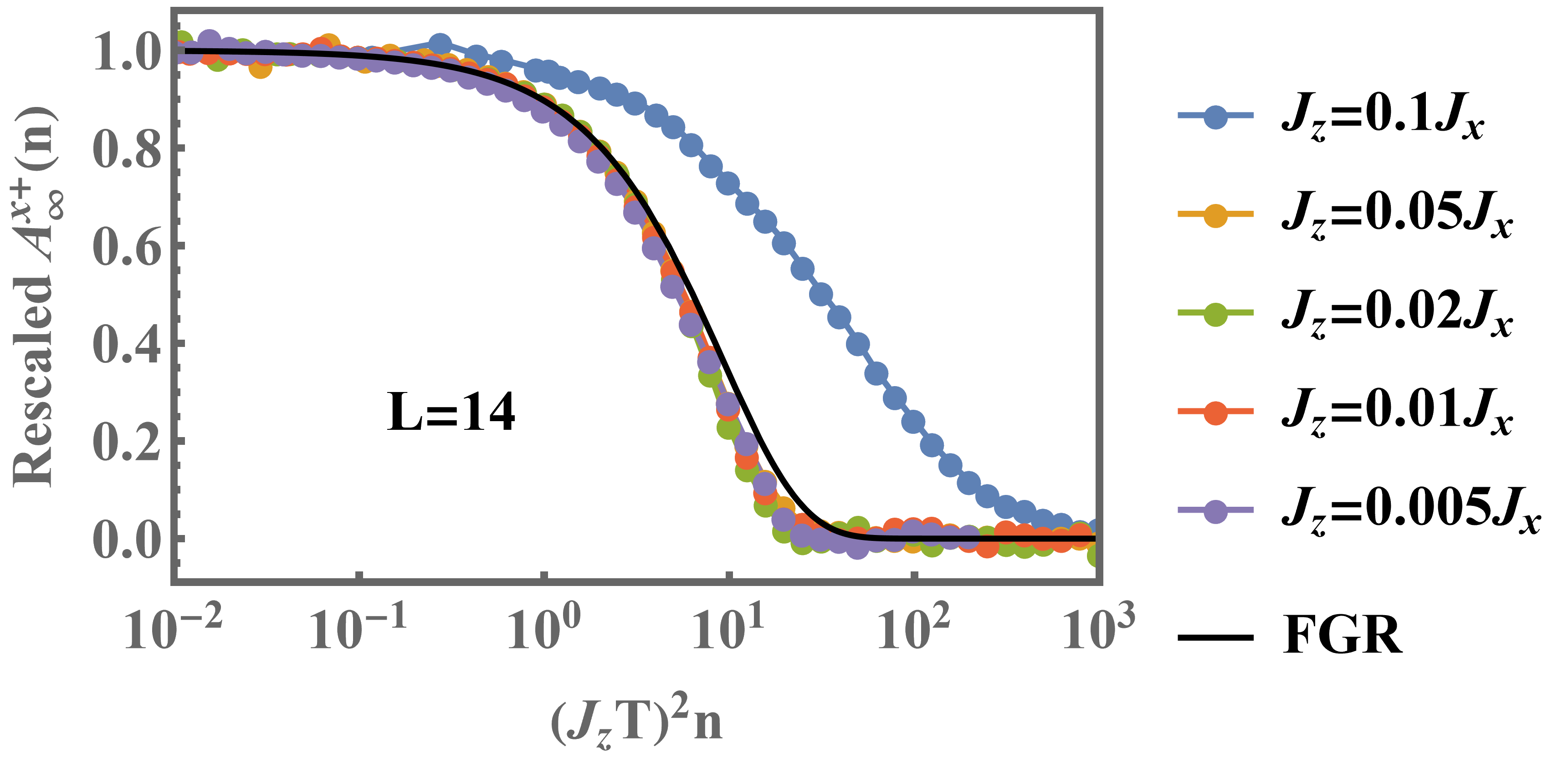}
    \includegraphics[width=0.32\textwidth]{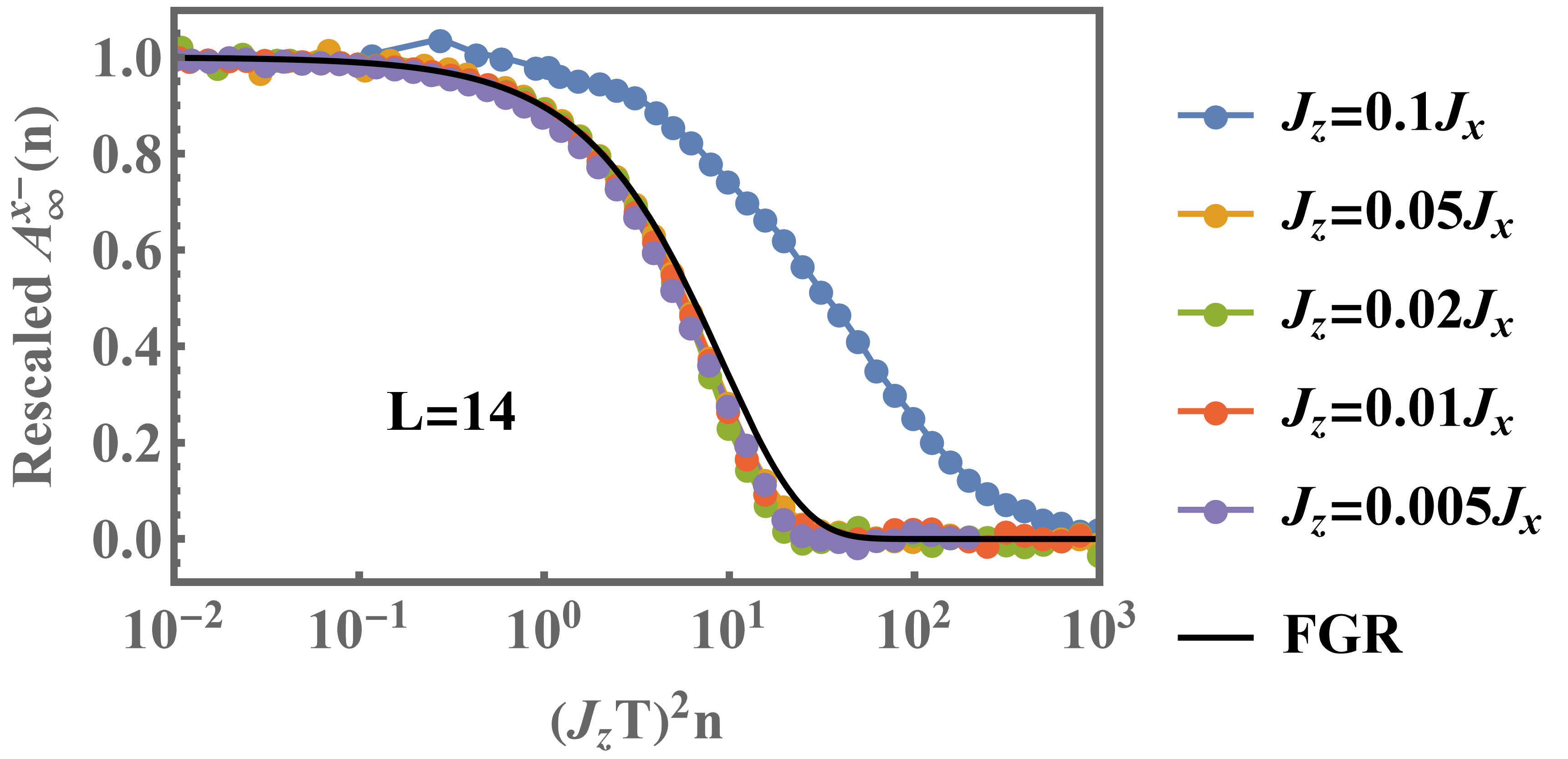}
    \includegraphics[width=0.32\textwidth]{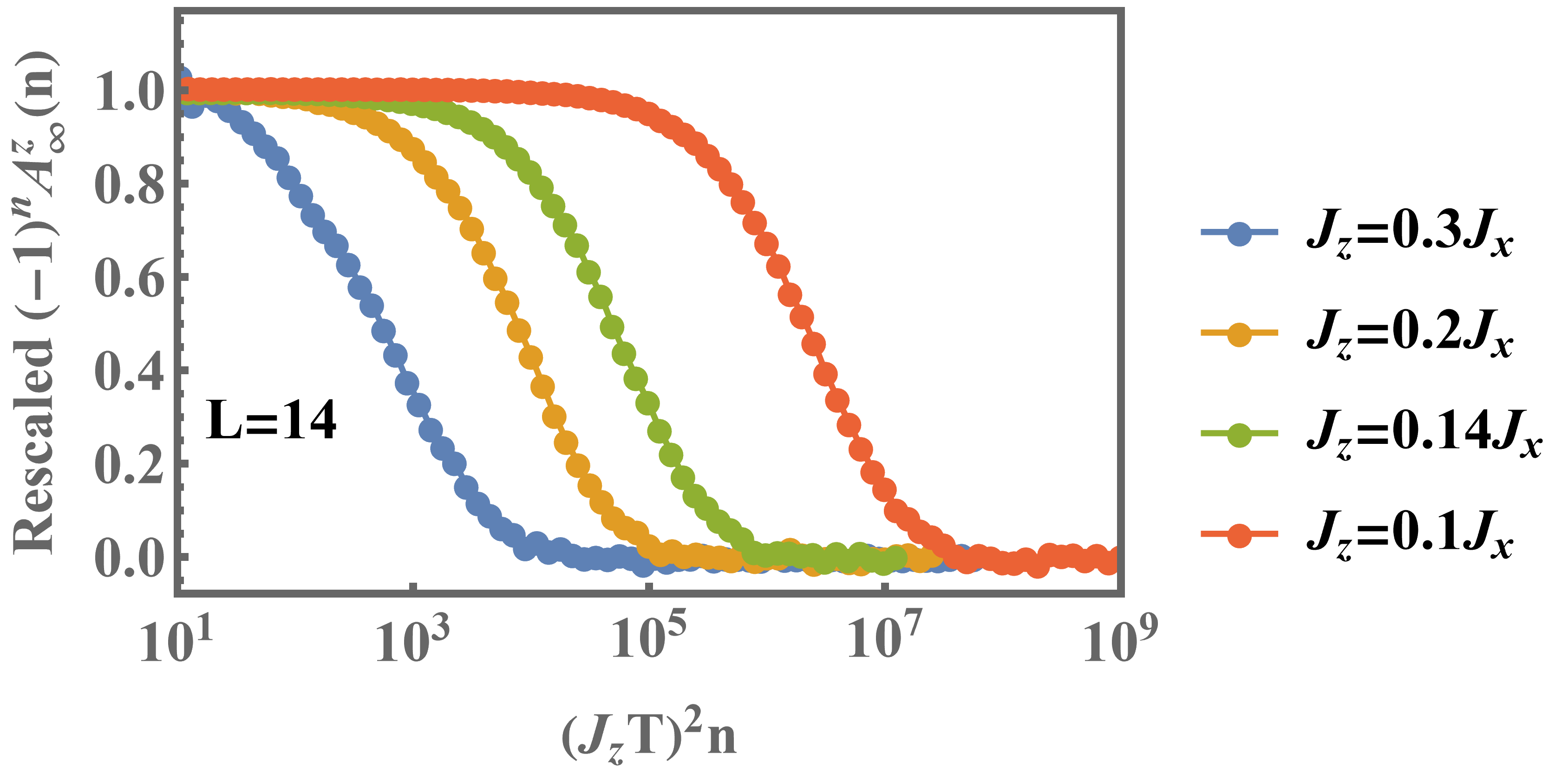}
    \caption{
    The autocorrelations for different strengths of the integrability breaking term $J_z$, and for $J_xT=2.8,gT=1.6$ (blue cross in Figures \ref{Fig: Phase Diagram}, \ref{Fig: Decay Channel}). Top panels: The autocorrelation for system sizes $L = 12,14$ for almost strong $0$-mode (left), almost strong $\pi$-mode (middle) and the product mode (right). The lifetime increases as $J_z$ decreases. For $J_z=0.1\,J_x$, the lifetime of the product mode is more than 4 orders of magnitude larger than the lifetime of the $0$ or $\pi$ modes.
    Bottom panels: The autocorrelation function is rescaled to be $1$ in the quasi-stable region. The time is rescaled to $(J_zT)^{2}n$ (all panels) and the autocorrelations is compared with the FGR result $\Gamma = 0.109 J_z^2 T$ (left and middle panels) because the scattering is dominated by the $\psi_0$-$\psi_\pi$ channel. The lifetime of the product mode (right) is much longer than almost strong $0$ and $\pi$-modes and the rescaled autocorrelation suggests a decay rate beyond FGR. Due to strong system size effects, the numerical results only probe up to $J_z = 0.1J_x$ and therefore we cannot conclude the scaling of the the autocorrelations for small $J_z/J_x\ll 1$.}
    \label{Fig: Autocorrelation}
\end{figure*}

In summary, the FGR decay rates for 0, $\pi$ and product modes are the sum of the following different channels, see details in Appendix \ref{Sec:A},
\begin{align}
    \Gamma_0 = \Gamma_{0,\text{B}} + \Gamma_{0,\pi};~
    \Gamma_\pi = \Gamma_{\pi,\text{B}} + \Gamma_{\pi,0};~
    \Gamma_\Psi = \Gamma_{0,\text{B}} + \Gamma_{\pi,\text{B}}.
    \label{Eq: FGR different channels}
\end{align}
Above $\Gamma_{0(\pi),\text{B}}$ is the FGR decay of the $0(\pi)$ mode through three bulk quasi-particles, while $\Gamma_{0,\pi} (\Gamma_{\pi,0})$ is the decay of the $0 (\pi)$ mode due to scattering with $\pi(0)$ edge mode and two bulk quasi-particles. 
The FGR decay of $0$-mode ($\pi$-mode) consists of $\psi_0$-bulk ($\psi_\pi$-bulk) and $\psi_0$-$\psi_\pi$ scattering, while the FGR decay of product mode consists of $\psi_0$-bulk and $\psi_\pi$-bulk scattering only. Interestingly, within FGR $\Gamma_{0,\pi}=\Gamma_{\pi,0}$, see Appendix \ref{Sec:A}.

There are parameters where FGR is no longer valid, needing one to perform higher order perturbation theory. Here we simply provide a simple counting argument for quasi-energy conservation that allows us to determine the leading power of $V$ controlling the decay.  For this we need the bulk dispersion of the unperturbed Floquet unitary $U_0$ with periodic boundary conditions \cite{yeh2023decay}
\begin{align}
    \cos(\epsilon_k T) =  \cos(gT)\cos(J_xT) + \sin(gT)\sin(J_x T)\cos{k}.
    \label{Eq: Bulk Dispersion}
\end{align}
Since the perturbation $V$ is a four-Majorana interaction, there are $4n$ Majoranas in $2n$-th order perturbation. Note that the decay rate should stay positive when the perturbation coupling flips sign, and therefore only even order of the perturbation contributes to the decay rate. We consider three cases: (i) $0$-mode scatters with $4n-1$ bulk excitations  corresponding to $\psi_0$-bulk channel, (ii) $\pi$-mode scatters with $4n-1$ bulk excitations corresponding to $\psi_\pi$-bulk channel, and (iii) $0$ and $\pi$ modes scatter with each other and $4n-2$ bulk excitations, corresponding to  $\psi_0$-$\psi_\pi$ channel. These correspond respectively to the following quasi-energy conservation conditions (modulo $2 \pi/{T}$)
\begin{align}
0=\sum_{i=1}^{4n-1} (\pm \epsilon_{k_i});
\pi/T=\sum_{i=1}^{4n-1} (\pm \epsilon_{k_i});
\pi/T=\sum_{i=1}^{4n-2} (\pm \epsilon_{k_i}), \label{Eq: Quasi-energy conservation}
\end{align}
where $\epsilon_{k_i}$ is the single particle bulk dispersion \eqref{Eq: Bulk Dispersion} and $\pm$ signs reflect that the scattering can involve either creation or annihilation of a bulk excitation. In Fig.~\ref{Fig: Decay Channel}, the colors show the lowest power of the perturbation controlling the decay rate, as determined by quasi-energy conservation \eqref{Eq: Quasi-energy conservation}. There exists a region where the product mode survives much longer than $0$ and $\pi$-modes, e.g., the blue cross in Fig.~\ref{Fig: Decay Channel}, where $\psi_0$-$\psi_\pi$ channel is 2nd order while the other channels are 4th order. This suggests the higher order decay rate of the product mode due to \eqref{Eq: FGR different channels}. In addition, even in the region where all channels are 2nd order, the product mode might have a longer lifetime compared to 0 and $\pi$-modes since the decay rate also depends on the localization lengths $\xi_0$ and $\xi_\pi$. Therefore, below we also study the dynamics keeping $\xi_0$ or $\xi_\pi$ fixed, which are the four dashed curves with arrows in Fig.~\ref{Fig: Decay Channel}.

\section{Results and Discussion}
\label{Sec: IV}
The autocorrelation of $\sigma_1^x$ detects the decay of 0 and $\pi$-mode at the same time. In the non-perturbed case, the autocorrelation obeys $A^x_\infty(n) \sim c_1 +c_2(-1)^n$, where $c_1$ and $c_2$ are  constants. The 0 ($\pi$)-mode contributes to $c_1 (c_2)$. Even in the presence of perturbations, this leads to two pronounced peaks in the spectral function  $\Tilde{A}_\infty^x(\omega) = \sum_n A_\infty^x(n)e^{-i\omega nT}$ at $\omega=0$ and $\omega=\pi/T$, see discussion in Appendix \ref{Sec:C}. 
The decay of the $0$ and $\pi$ modes in $n$ can be studied directly through the following decomposition
\begin{align}
    &A_\infty^{x+}(n+1/2) = \frac{A_\infty^x (n+1)+A_\infty^x (n)}{2};\label{Eq: A+}\\
    &A_\infty^{x-}(n+1/2) = \frac{A_\infty^x (n+1) -A_\infty^x (n)}{2}.\label{Eq: A-}
\end{align} 
We utilize the data points where $n$ is odd in \eqref{Eq: A+} and \eqref{Eq: A-} ensuring a positive sign for $A^{x-}_{\infty}$. The parameters are chosen to be $J_xT =2.8$ and $gT = 1.6$, corresponding to the blue cross in Figures \ref{Fig: Phase Diagram}, \ref{Fig: Decay Channel}. The results $A^{\pm}_{\infty}$ are shown in the left and middle panels of Fig.~\ref{Fig: Autocorrelation}. The right panels shows the autocorrelation  $(-1)^nA^{z}_{\infty}$. The bottom left and middle panels are rescaled plots that highlight the agreement with FGR for small $J_z$. Due to  strong finite size effects, we cannot probe small $J_z/J_x \ll 1$ for the decay of the product mode. However the plots already show a much longer lifetime for the product mode, even for the larger values of $J_z$, as compared to $0,\pi$ modes in Fig.~\ref{Fig: Autocorrelation}. This is consistent with the quasi-energy conservation calculation in Fig.~\ref{Fig: Decay Channel}.

\begin{figure}[h!]
    \centering
    \includegraphics[width=0.45\textwidth]{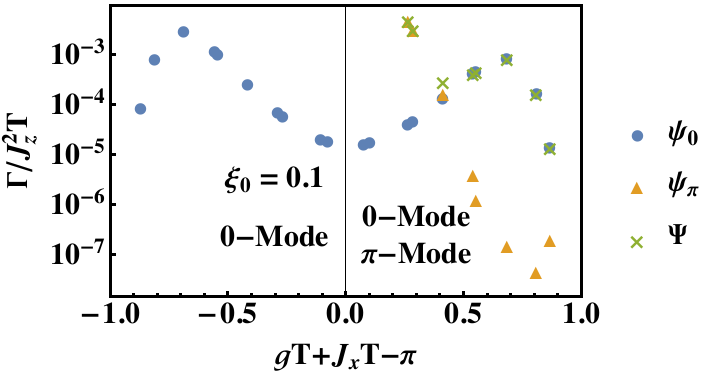}
    \includegraphics[width=0.45\textwidth]{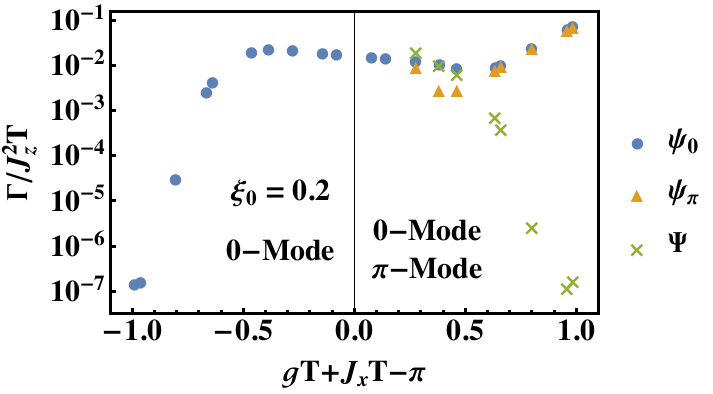}
    \caption{FGR results for fixed localization length of the $0$-mode, $\xi_0 = 0.1$ (top) and $\xi_0 = 0.2$ (bottom) for $L=50$, and corresponding respectively to the upper and lower left curves in Fig.~\ref{Fig: Decay Channel}. Due to rapid oscillations at long times, the numerical accuracy of the FGR result is $\approx 10^{-6}$. For $\xi_0 = 0.1$ (top), the FGR decay only arises from $\psi_0$-bulk and $\psi_\pi$-bulk channels, with the decay rate of the product mode being the sum of $0$ and $\pi$-mode decay rate according to \eqref{Eq: FGR different channels}. The asymmetry of the $0$-mode decay rate indicates the suppression of the matrix element due to the presence of the $\pi$-mode. The decay rate of $\pi$-mode is suppressed faster, reflecting the approach to the higher order region in Fig.~\ref{Fig: Decay Channel}.  For $\xi_0 = 0.2$ (lower-panel), all scattering channels contribute in FGR. However, the matrix element is dominated by $\psi_0$-$\psi_\pi$ channel as the $\pi$-mode  is localized. Therefore, the decay rate of $0$ and $\pi$-mode are the same for the data points on the right. The product mode has a much smaller decay rate on the right because it only decays via scattering with bulk excitations, see \eqref{Eq: FGR different channels}.}
    \label{Fig: FGR 0-mode fixed}
\end{figure}

We also perform the FGR calculation along the four curves in 
Figures \ref{Fig: Phase Diagram},\ref{Fig: Decay Channel}.  The FGR results of the left curves (fixed $\xi_0$) are shown in Fig.~\ref{Fig: FGR 0-mode fixed}. The top panels in Fig.~\ref{Fig: FGR 0-mode fixed} explore the decay rate when it is dominated by bulk channels such that $\Gamma_\Psi = \Gamma_0 + \Gamma_\pi$. From \eqref{Eq: Bulk Dispersion}, the bulk dispersion is invariant under reflection about $gT+J_xT=\pi$ and $gT = J_xT$, and the $x$-axes reflect this. The asymmetry in the decay rate of the $0$-mode in the top panel of Fig.~\ref{Fig: FGR 0-mode fixed} indicates that the existence of $\pi$-mode suppresses the $\psi_0$-bulk channel. This can be understood in the limiting case of completely localized edge modes, $\psi_0 = \alpha_1$ and $\psi_\pi = \beta_1$. The perturbation now is purely in the  $\psi_0$-$\psi_\pi$ channel, $\psi_0\psi_\pi\Tilde{\alpha}_2\Tilde{\beta}_2$, and bulk channels are fully suppressed. 
In general, although the bulk channels are the leading contributions in the top panels of Fig.~\ref{Fig: FGR 0-mode fixed}, it is suppressed for the right data points, because of the localized edge modes. This effect also appears in the bottom panels of Fig.~\ref{Fig: FGR 0-mode fixed}, where all three channels obey FGR, and yet, the product model is more robust than both $0$ and $\pi$-modes since it only senses the bulk channels. 

The results for the right curves in Figures \ref{Fig: Phase Diagram}, \ref{Fig: Decay Channel} (fixed $\xi_\pi$) share the same features as Fig.~\ref{Fig: FGR 0-mode fixed} and are presented in Appendix \ref{Sec:B}, where the details of the numerical computation are also discussed. Due to oscillations in time, decay rates of $O(10^{-6})$ and smaller are not accurate, yet these data points are presented in order to indicate a slow decay channel.

\section{Conclusions}
\label{Sec: V}
We have presented a new mode, dubbed the Floquet product mode, that arises from the product of two existing modes, but nevertheless has dynamics which is independent of the modes from which it is constructed. We have shown that
the product mode can have parametrically longer lifetimes than the constituent modes.
This result can be interpreted as a Majorana analog of an approximate decoherence free subspace, but generalized to an interacting and infinite temperature system. From the viewpoint of diagrammatic perturbation theory, the lifetime of the constituent Majorana states is encoded in the imaginary part of the Green's function, with the long
lifetime of the product mode reflecting a subtle cancellation between vertex- and self-energy corrections.

While we presented this model for a non-integrable Floquet Ising chain, 
we expect that similar effects occur in periodically driven topological superconductors \cite{Matthies2022}. 
The model investigated by us can be realized in a straightforward way on present-day quantum computers, however, additional decay channels arising from noise and dephasing effects will play a role in this case.

\emph{Acknowledgments:} This work was supported by the US Department of Energy, Office of
Science, Basic Energy Sciences, under Award No.~DE-SC0010821 (HY, AM) and by the  German Research Foundation within
CRC183 (project number 277101999, subproject A01 (AR)). HY acknowledges support of the NYU IT High Performance Computing resources, services, and staff expertise.

\appendix

\begin{widetext}
\section{Decay channels in FGR}
\label{Sec:A}
Here we provide a detailed derivation of the different decay channels of the edge modes within a Fermi Golden Rule (FGR) approximation. In Ref.~\cite{yeh2023decay}, the FGR decay rate when only a $0$ or $\pi$-mode is present, was derived. These were shown to be 
\begin{align}
    &\Gamma_0 = \frac{T}{2^L} 
    \biggl( \frac{1}{2}  \text{Tr} \left[ \dot{\psi_0} \dot{\psi_0} \right] + \sum_{n= 1}^{\infty} \text{Tr}\left[\dot{\psi_0}(n) \dot{\psi_0}\right] \biggr)= \frac{1}{2^L} \sum_{i,j} |\langle i| \Dot{\psi_0} |j\rangle|^2  \pi \delta_F\left(\epsilon_i -\epsilon_j \right);\label{Eq: FGR of 0 edge mode}\\
    &\Gamma_\pi = \frac{T}{2^L} 
    \biggl( \frac{1}{2}  \text{Tr} \left[ \dot{\psi_\pi} \dot{\psi_\pi} \right] + \sum_{n= 1}^{\infty} (-1)^n\text{Tr}\left[\dot{\psi_\pi}(n) \dot{\psi_\pi}\right] \biggr)= \frac{1}{2^L} \sum_{i,j} |\langle i| \Dot{\psi_\pi} |j\rangle|^2  \pi \delta_F\left(\epsilon_i -\epsilon_j + \frac{\pi}{T} \right),
    \label{Eq: FGR of pi edge mode}
\end{align}
where $\Dot{\psi}_{0/\pi} = i[V,\psi_{0/\pi}]$ and $V = J_z H_{zz}/2$. We define $\Dot{\psi}_{0/\pi} (n) = (U_0^\dagger)^n \Dot{\psi}_{0/\pi} U_0^n$ with the unperturbed Floquet unitary $U_0$. $| i \rangle$ are the many-particle eigenstates of $U_0$ with eigenvalue $e^{-i \epsilon_i T}$ and the $\delta_F$ function encodes energy conservation modulo $2 \pi/{T}$, with $\delta_F(\epsilon)=\sum_{m\in {\rm int}} \delta(\epsilon+m 2\pi/T)$. The analytic expression of edge modes are \cite{yeh2023decay}
\begin{align}
    &\psi_0 = N_0 \sum_{l = 1}^\infty\left[ \cos\left(\frac{gT}{2}\right)a_l + \sin\left(\frac{gT}{2}\right)b_l\right] \xi_0^{l-1}; 
    &\psi_\pi = N_\pi \sum_{l = 1}^\infty\left[ \sin\left(\frac{gT}{2}\right)a_l - \cos\left(\frac{gT}{2}\right)b_l\right] \xi_\pi^{l-1}, 
\end{align}
where $\mathcal{N}_0$ and $\mathcal{N}_\pi$ are normalization prefactors and the localization length of edge modes are: $\xi_0 = \tan(gT/2)\cot(J_xT/2)$ and $\xi_\pi = -\cot(gT/2)\cot(J_xT/2)$. The Majoranas are defined according to the following convention
\begin{align}
    &a_{l} = \prod_{j=1}^{l-1}\sigma_j^z \sigma_l^x; & b_{l} = \prod_{j=1}^{l-1}\sigma_j^z \sigma_l^y.
\end{align}
By rotation of the basis, we define the Majoranas as
\begin{align}
    &\alpha_l = \cos\left(\frac{gT}{2}\right)a_l + \sin\left(\frac{gT}{2}\right)b_l;
    &\beta_l = \sin\left(\frac{gT}{2}\right)a_l - \cos\left(\frac{gT}{2}\right)b_l.
\end{align}
In the new basis, the edge modes and the perturbation $V$ have simple expressions
\begin{align}
    &\psi_0 = N_0 \sum_{l = 1}^\infty\alpha_l \xi_0^{l-1}; \quad\quad\quad\quad\quad\quad
    \psi_\pi = N_\pi \sum_{l = 1}^\infty\beta_l \xi_\pi^{l-1};
    & V =  -\frac{J_z}{2}\sum_l^\infty \alpha_l \beta_l \alpha_{l+1} \beta_{l+1}.
\end{align}
Now, one can proceed to calculate FGR \eqref{Eq: FGR of 0 edge mode} and \eqref{Eq: FGR of pi edge mode} with the above expressions. 

The Majoranas $\{\alpha_l, \beta_l \}$ consist of edge and bulk degrees of freedom, i.e., $\alpha_l = (\psi_0|\alpha_l)\psi_0 + \Tilde{\alpha}_l$ and $\beta_l = (\psi_\pi|\beta_l)\psi_\pi + \Tilde{\beta}_l$ with the inner product between two operator defined as $(A|B) = \text{Tr}[A^\dagger  B]/2^L$. Let us first focus on the $0$-mode. The perturbation can be separated into terms commuting or non-commuting with the $0$-mode by employing $\alpha_l = (\psi_0|\alpha_l)\psi_0 + \Tilde{\alpha}_l$,
\begin{align}
    V = -\frac{J_z}{2}\sum_{l=1}^{\infty} \left[(\psi_0|\alpha_l)\times\psi_0 \beta_l \Tilde{\alpha}_{l+1}\beta_{l+1} + (\psi_0|\alpha_{l+1})\times\Tilde{\alpha}_l \beta_l \psi_0 \beta_{l+1}\right] + \text{terms commuting with $\psi_0$}
\end{align}
Only terms containing one $\psi_0$ lead to $[V,\psi_0] \neq 0$. Therefore, the $\Dot{\psi}_0$ term in the FGR decay rate is given by
\begin{align}
    \Dot{\psi_0} = i\left[V,\psi_0\right] =iJ_z \sum_{l=1}^\infty \left[ (\psi_0|\alpha_l) \times  \beta_{l} \Tilde{\alpha}_{l+1}\beta_{l+1} + (\psi_0|\alpha_{l+1}) \times
    \Tilde{\alpha}_{l} \beta_{l} \beta_{l+1} \right].
\end{align}
Moreover, using that the $\beta$ Majoranas have an overlap with the $\pi$ mode and the bulk modes, $\beta_l = (\psi_\pi|\beta_l)\psi_\pi + \Tilde{\beta}_l$, the above expression can be further expanded into the sum of two channels $\Dot{\psi_0} = \Dot{\psi_0}_{,\pi}+  \Dot{\psi_0}_{,\text{B}}$, where
\begin{align}
   \Dot{\psi_0}_{,\pi} = iJ_z \sum_{l=1}^\infty \Big[ &(\psi_0|\alpha_l)(\psi_\pi|\beta_l) \times \psi_\pi \Tilde{\alpha}_{l+1} \Tilde{\beta}_{l+1} + (\psi_0|\alpha_l)(\psi_\pi|\beta_{l+1}) \times  \Tilde{\beta}_{l} \Tilde{\alpha}_{l+1} \psi_\pi \nonumber\\
   &+(\psi_\pi|\beta_l)(\psi_0|\alpha_{l+1}) \times \Tilde{\alpha}_{l}\psi_\pi\Tilde{\beta}_{l+1} + (\psi_0|\alpha_{l+1})(\psi_\pi|\beta_{l+1}) \times 
    \Tilde{\alpha}_{l} \Tilde{\beta}_{l}\psi_\pi\Big],\label{Eq: psidot 0-pi}
    \\
    \Dot{\psi_0}_{,\text{B}} = iJ_z \sum_{l=1}^\infty \Big[& (\psi_0|\alpha_l) \times  \Tilde{\beta}_{l} \Tilde{\alpha}_{l+1}\Tilde{\beta}_{l+1} + (\psi_0|\alpha_{l+1}) \times
    \Tilde{\alpha}_{l} \Tilde{\beta}_{l} \Tilde{\beta}_{l+1} \nonumber\\
    &-(\psi_0|\alpha_l)(\psi_\pi|\beta_l)(\psi_\pi|\beta_{l+1}) \times \Tilde{\alpha}_{l+1} + (\psi_\pi|\beta_l)(\psi_0|\alpha_{l+1})(\psi_\pi|\beta_{l+1}) \times \Tilde{\alpha}_{l}\Big].
\end{align}
Above, $\Dot{\psi_0}_{,\pi}$ describes scattering between the $0$ and $\pi$-modes and $\Dot{\psi_0}_{,\text{B}}$ accounts for scattering of the $0$ mode with the bulk modes. Substituting the above expressions in \eqref{Eq: FGR of 0 edge mode} and \eqref{Eq: FGR of pi edge mode}, the FGR decay rate of the zero mode now consists of two parts, $\Gamma_0 = \Gamma_{0,\pi} + \Gamma_{0,\text{B}}$,
\begin{align}
    &\Gamma_{0,\pi} = \frac{T}{2^L} 
    \biggl( \frac{1}{2}  \text{Tr} \left[ \Dot{\psi_0}_{,\pi} \Dot{\psi_0}_{,\pi} \right] + \sum_{n= 1}^{\infty} \text{Tr}\left[\Dot{\psi_0}_{,\pi}(n) \Dot{\psi_0}_{,\pi}\right] \biggr) = \frac{1}{2^L} \sum_{i,j} |\langle i| \Dot{\psi_0}_{,\pi} |j\rangle|^2  \pi \delta_F\left(\epsilon_i -\epsilon_j \right),\\
    &\Gamma_{0,\text{B}} = \frac{T}{2^L} 
    \biggl( \frac{1}{2}  \text{Tr} \left[ \Dot{\psi_0}_{,\text{B}} \Dot{\psi_0}_{,\text{B}} \right] + \sum_{n= 1}^{\infty} \text{Tr}\left[\Dot{\psi_0}_{,\text{B}}(n) \Dot{\psi_0}_{,\text{B}}\right] \biggr) = \frac{1}{2^L} \sum_{i,j} |\langle i| \Dot{\psi_0}_{,\text{B}} |j\rangle|^2  \pi \delta_F\left(\epsilon_i -\epsilon_j \right).\label{Eq: FGR of 0-bulk}
\end{align}
Note that the cross term $\text{Tr}[\Dot{\psi_0}_{,\pi}\Dot{\psi_0}_{,\text{B}}] = 0$ since $\text{Tr}[\psi_\pi] = 0$ and only a single $\psi_\pi$ enters inside the trace of the cross term.

Similarly, for the case of $\pi$-mode, one expands the perturbation with $\beta_l = (\psi_\pi|\beta_l)\psi_\pi + \Tilde{\beta}_l$ and derives $\Dot{\psi_\pi}$ as
\begin{align}   \Dot{\psi_\pi} =   i\left[V,\psi_\pi\right] = -iJ_z \sum_{l=1}^\infty \Big[& (\psi_\pi|\beta_l) \times  \alpha_l \alpha_{l+1}\Tilde{\beta}_{l+1}  + (\psi_\pi|\beta_{l+1}) \times
    \alpha_l \Tilde{\beta}_l \alpha_{l+1} \Big].
\end{align}
The above can be further separated into two channels by using $\alpha_l = (\psi_0|\alpha_l)\psi_0 + \Tilde{\alpha}_l$,
\begin{align}
   \Dot{\psi_\pi}_{,0} = -iJ_z \sum_{l=1}^\infty \Big[ & (\psi_\pi|\beta_l)(\psi_0|\alpha_l) \times  \psi_0 \Tilde{\alpha}_{l+1}\Tilde{\beta}_{l+1}+ (\psi_\pi|\beta_l)(\psi_0|\alpha_{l+1}) \times  \Tilde{\alpha_l} \psi_0\Tilde{\beta}_{l+1}\nonumber\\
   &+(\psi_\pi|\beta_{l+1})(\psi_0|\alpha_l) \times
    \psi_0 \Tilde{\beta}_l \Tilde{\alpha}_{l+1} + (\psi_\pi|\beta_{l+1})(\psi_0|\alpha_{l+1}) \times
    \Tilde{\alpha_l} \Tilde{\beta}_l \psi_0\Big],\label{Eq: psidot pi-0}
    \\
    \Dot{\psi_\pi}_{,\text{B}} = -iJ_z \sum_{l=1}^\infty \Big[& (\psi_\pi|\beta_l) \times  \Tilde{\alpha}_l \Tilde{\alpha}_{l+1}\Tilde{\beta}_{l+1}  + (\psi_\pi|\beta_{l+1}) \times
    \Tilde{\alpha}_l \Tilde{\beta}_l \Tilde{\alpha}_{l+1} \nonumber\\
    &+(\psi_\pi|\beta_l)(\psi_0|\alpha_l)(\psi_0|\alpha_{l+1}) \times  \Tilde{\beta}_{l+1}  - (\psi_\pi|\beta_{l+1})(\psi_0|\alpha_l)(\psi_0|\alpha_{l+1}) \times
    \Tilde{\beta}_l \Big].
\end{align}
Hence, the FGR is the sum of two channels, $\Gamma_\pi = \Gamma_{\pi,0} + \Gamma_{\pi,\text{B}}$,
\begin{align}
    &\Gamma_{\pi,0} = \frac{T}{2^L} 
    \biggl( \frac{1}{2}  \text{Tr} \left[ \Dot{\psi_\pi}_{,0} \Dot{\psi_\pi}_{,0} \right] + \sum_{n= 1}^{\infty} (-1)^n\text{Tr}\left[\Dot{\psi_\pi}_{,0}(n) \Dot{\psi_\pi}_{,0}\right] \biggr) = \frac{1}{2^L} \sum_{i,j} |\langle i| \Dot{\psi_\pi}_{,0} |j\rangle|^2  \pi \delta_F\left(\epsilon_i -\epsilon_j + \frac{\pi}{T} \right),\\
    &\Gamma_{\pi,\text{B}} = \frac{T}{2^L} 
    \biggl( \frac{1}{2}  \text{Tr} \left[ \Dot{\psi_\pi}_{,\text{B}} \Dot{\psi_\pi}_{,\text{B}} \right] + \sum_{n= 1}^{\infty} (-1)^n\text{Tr}\left[\Dot{\psi_\pi}_{,\text{B}}(n) \Dot{\psi_\pi}_{,\text{B}}\right] \biggr) = \frac{1}{2^L} \sum_{i,j} |\langle i| \Dot{\psi_\pi}_{,\text{B}} |j\rangle|^2  \pi \delta_F\left(\epsilon_i -\epsilon_j + \frac{\pi}{T} \right).
\end{align}

Finally, we consider the FGR of the product mode, $\Psi = i\psi_0\psi_\pi$. Since $\Psi$ behaves like a $\pi$ mode, $U_0^\dagger\Psi U_0 = - \Psi$, the FGR of $\Psi$ obeys 
\begin{align}
    \Gamma_\Psi &=  \frac{T}{2^L} 
    \biggl( \frac{1}{2}  \text{Tr} \left[ \dot{\Psi} \dot{\Psi} \right] + \sum_{n= 1}^{\infty} (-1)^n\text{Tr}\left[\dot{\Psi}(n) \dot{\Psi}\right] \biggr) = \frac{1}{2^L} \sum_{i,j} |\langle i| \Dot{\Psi} |j\rangle|^2  \pi \delta_F\left(\epsilon_i -\epsilon_j + \frac{\pi}{T} \right).
\end{align}
Using chain rule, $\Dot{\Psi}$ can be expressed as
\begin{align}\label{chainR}
    \Dot{\Psi} = i\Dot{\psi_0} \psi_\pi + i\psi_0 \Dot{\psi_\pi} = i\Dot{\psi_0}_{,\text{B}} \psi_\pi + i\psi_0 \Dot{\psi_\pi}_{,\text{B}} + i(\Dot{\psi_0}_{,\pi}\psi_\pi + \psi_0 \Dot{\psi_\pi}_{,0}).
\end{align}
Note that 
\begin{align}
\Dot{\psi_0}_{,\pi}\psi_\pi + \psi_0 \Dot{\psi_\pi}_{,0} = 0, \label{id1}
\end{align} which can be checked directly from \eqref{Eq: psidot 0-pi} and \eqref{Eq: psidot pi-0}. Here we provide a simple argument. Since the commutator $[\psi_0\psi_\pi, \psi_0\psi_\pi \Tilde{\alpha}\Tilde{\beta}] = 0$ forbidding the scattering between $0$ and $\pi$-modes, it implies that non-zero commutations only come from the bulk channels. Therefore, the FGR of the product mode can be expressed as
\begin{align}
    \Gamma_\Psi &= \frac{T}{2^L} 
    \biggl\{ \frac{1}{2}  \Big(\text{Tr}[\Dot{\psi_0}_{,\text{B}} \Dot{\psi_0}_{,\text{B}}] + \text{Tr}[\Dot{\psi_\pi}_{,\text{B}}\Dot{\psi_\pi}_{,\text{B}}]\Big)+ \sum_{n= 1}^{\infty} \Big(\text{Tr}[\Dot{\psi_0}_{,\text{B}} (n)\Dot{\psi_0}_{,\text{B}}] + (-1)^n\text{Tr}[\Dot{\psi_\pi}_{,\text{B}}(n)\Dot{\psi_\pi}_{,\text{B}}]\Big)\biggr\} \nonumber\\
    &= \Gamma_{0,\text{B}} + \Gamma_{\pi,\text{B}}
\end{align}
Note that the cross terms are not allowed since trace of odd numbers of $\psi_0$ or $\psi_\pi$ is zero.
In summary, the FGR decay rates for the  $0$, $\pi$ and product modes are given by
\begin{align}
    &\Gamma_0 = \Gamma_{0,\text{B}} + \Gamma_{0,\pi};\quad\quad\quad\quad\quad\quad
    \Gamma_\pi = \Gamma_{\pi,\text{B}} + \Gamma_{\pi,0};
    &\Gamma_\Psi = \Gamma_{0,\text{B}} + \Gamma_{\pi,\text{B}}.
\end{align}

We now show that $\Gamma_{0,\pi} = \Gamma_{\pi,0}$ at second order in the perturbation. We use the identity \eqref{id1} in the
FGR formula,
\begin{align}
    \Gamma_{\pi,0} &= \frac{1}{2^L} \sum_{i,j}|\langle i|\Dot{\psi_{\pi}}_{,0}|j\rangle|^2 \pi \delta_F\left(\epsilon_i -\epsilon_j + \frac{\pi}{T} \right) \nonumber\\
    &= \frac{1}{2^L} \sum_{i,j}|\langle i|\psi_0\Dot{\psi_{0}}_{,\pi}\psi_\pi|j\rangle|^2 \pi \delta_F\left(\epsilon_i -\epsilon_j + \frac{\pi}{T} \right),\ \text{by } \Dot{\psi_0}_{,\pi}\psi_\pi + \psi_0 \Dot{\psi_\pi}_{,0} = 0\nonumber\\
    &= \frac{1}{2^L} \sum_{k,l}|\langle k|\Dot{\psi_{0}}_{,\pi}|l\rangle|^2 \pi \delta_F\left(\epsilon_k -\epsilon_l \right) = \Gamma_{0, \pi},
\end{align}
where in the last line we relabel states as $|k\rangle = \psi_0 |i\rangle$ and $|l\rangle =\psi_\pi|j\rangle$ with quasi-energy, $\epsilon_k = \epsilon_i$ and $\epsilon_l = \epsilon_j - \pi/T$.

In summary, the ``symmetry'' $\Gamma_{0,\pi} = \Gamma_{\pi,0}$ in second order is a consequence of the operator relation \eqref{id1}. For decay rates beyond second order, the meaning of $\Gamma_{0,\pi}$ may be a little ambiguous. For example, one may consider a higher order process for the $0$-mode self-energy bubble diagram where some of the internal lines are $\pi$-mode propagators that are not directly connected to $0$-mode propagators, i.e, $0$-mode and $\pi$-mode are scattered indirectly via bulk excitations. Therefore establishing $\Gamma_{0,\pi} = \Gamma_{\pi,0}$ in higher order processes is not well posed.
\end{widetext}

\section{Numerical Computation of FGR}
\label{Sec:B}

\begin{figure*}
    \centering
    \includegraphics[width=0.45\textwidth]{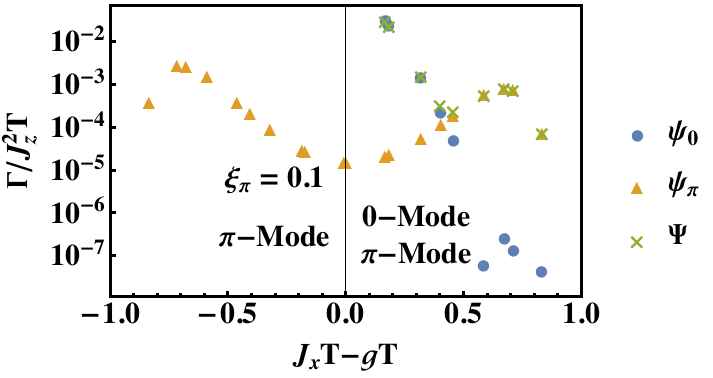}
    \includegraphics[width=0.45\textwidth]{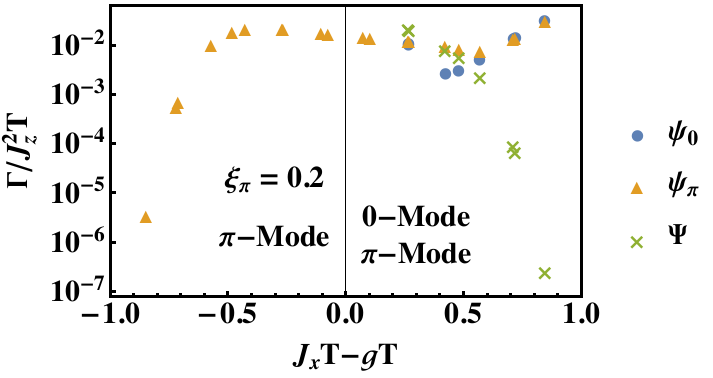}
    \caption{FGR results for fixed localization length of the $\pi$-mode, $\xi_\pi = 0.1$ (left) and $\xi_\pi = 0.2$ (right) for $L=50$. Due to rapid oscillations at long times, the numerical accuracy of FGR result is $\approx 10^{-6}$. Similar to the examples of fixed $\xi_0$ in Fig.~\ref{Fig: FGR 0-mode fixed}, FGR arises from $\psi_0$-bulk and $\psi_\pi$-bulk scattering channels for  $\xi_\pi = 0.1$ (left) while all channels are allowed in FGR for $\xi_\pi = 0.2$ (right).}
    \label{Fig: FGR pi-mode fixed}
\end{figure*}

\begin{figure*}
    \centering
    \includegraphics[width=0.45\textwidth]{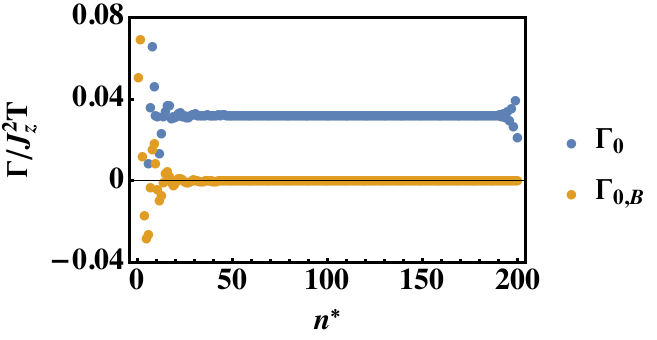}
    \includegraphics[width=0.45\textwidth]{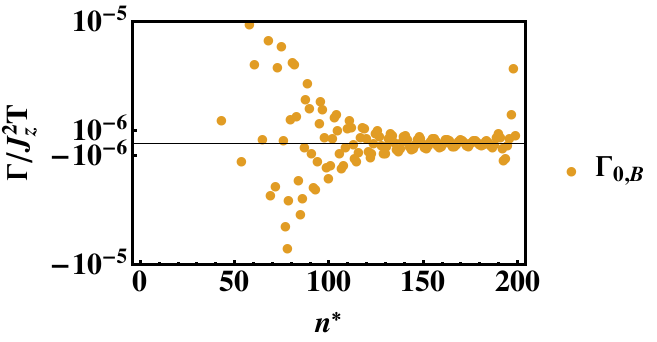}
    \caption{FGR results of $\Gamma_0$ and $\Gamma_{0,B}$ with $gT=1.8$ and $J_xT=2.65$, corresponding to the rightmost data in the right panel of Fig.~\ref{Fig: FGR pi-mode fixed}. Left panel: Numerical computation of decay rate by summing up to $n^*$ in \eqref{Eq: FGR of 0 edge mode} and \eqref{Eq: FGR of 0-bulk}. The late time oscillations are due to revivals in a finite size system. Right panel: $\Gamma_{0,B}$ shows oscillations with $n^*$ of order $10^{-6}$ between $n^* =110$ and $n^*=190$.}
    \label{Fig: Numerical Error}
\end{figure*}

In this section, we present the numerical method to perform FGR calculations for large system sizes. The FGR is related to the autocorrelation $\text{Tr}\left[\Dot{\psi}_{0/\pi}(n) \Dot{\psi}_{0/\pi}\right]$, and it leads to products of maximally six operators in the trace, e.g., $\text{Tr}\left[O_6 O_5 O_4 O_3 O_2 O_1\right]/2^L$. Each operator is a single Majorana evolving with the unperturbed Floquet unitary $U_0$. The latter is a matrix of size $2L \times 2L$, and is much easier to compute compared to a $2^L \times 2^L$ matrix in the many-particle basis. However, the matrix is represented in the single Majorana basis, and one cannot multiply and trace the matrices directly since multiplication and trace is defined on the many-particle basis. Instead, we will utilize the anticommutation property of Majoranas to calculate the trace. First, we perform the Gram-Schmidt orthogonalization of the six operators
\begin{align}
    &O_1 = c_1 \Bar{O}_1\\
    &O_i = \sum_{j=1}^{i-1}(\Bar{O}_j|O_i)\Bar{O}_j + c_i\Bar{O}_i,\ \text{for $i \geq 2$},
\end{align}
where we introduce coefficients $\{c_i\}$ such that $\{ \Bar{O}_i\}$ are normalized and $\{\Bar{O}_i, \Bar{O}_j \} = 2\delta_{ij}$. One can show that the following operators 
\begin{align}
    &O_1' = c_1 \Bar{O}_i'\\
    &O_i' = \sum_{j=1}^{i-1}(\Bar{O}_j|O_i)\Bar{O}_j' + c_i \Bar{O}_i',\ \text{for $i \geq 2$},
\end{align}
where $\{ \Bar{O}_i',\Bar{O}_j'\} =2\delta_{ij}$, leads to the same result for the trace, \newline $\text{Tr}\left[O_6' O_5' O_4' O_3' O_2' O_1'\right]/ D = \text{Tr}\left[O_6 O_5 O_4 O_3 O_2 O_1\right]/2^L$,  where $\{ O_i'\}$ are $D \times D$ matrices and $\{O_i \}$ are $2^L \times 2^L$ matrices. The precise value of $D$ is explained below. Inside the trace, only the algebra between $\{\Bar{O}_i\}$ matters and one can replace them by any other operators  $\{\Bar{O}_i'\}$ as long as the algebraic structure stays the same. Numerically, one first performs the Gram-Schmidt to obtain the coefficient $(\Bar{O}_j|O_i)$, and then computes the trace of the operator $O_i'$ by setting $\{ \Bar{O}_i' \} = \{ a_1, b_1, a_2, b_2, a_3, b_3 \}$, the first six Majoranas.  Since  the six Majoranas can be expressed as Pauli strings of a 1D spin chain on three sites, each of them is a $8\times 8$ matrix, thus $D=8$.

\begin{figure*}
    \centering
    \includegraphics[width=0.32\textwidth]{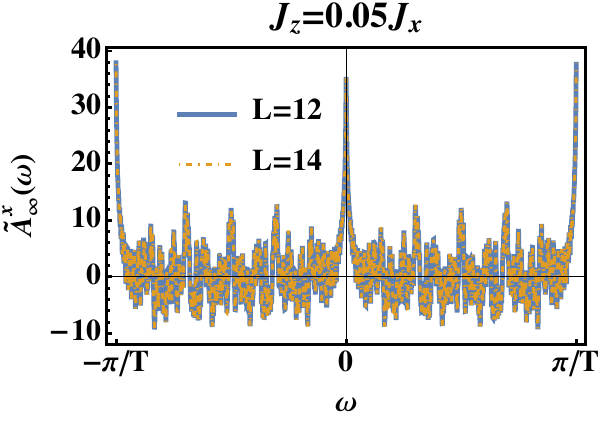}
    \includegraphics[width=0.32\textwidth]{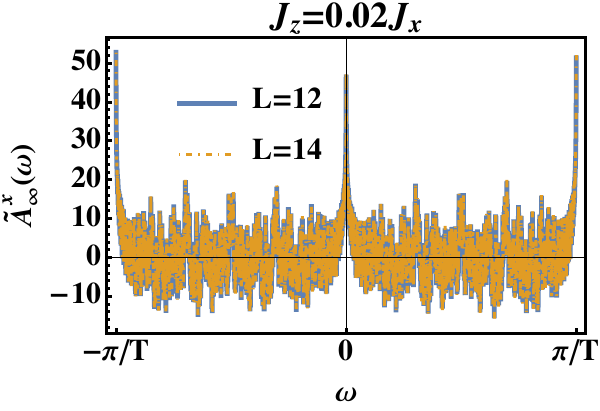}
    \includegraphics[width=0.32\textwidth]{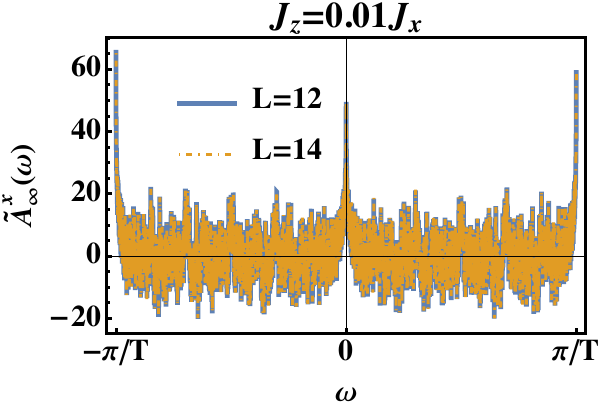}
    \caption{
    The spectral function $\Tilde{A}_\infty^x(\omega)$ for the same parameters as Fig.~\ref{Fig: Autocorrelation}. The peaks at $\omega = 0$ and $\pi/T$ indicate the almost strong $0$ and $\pi$-modes. The oscillations come from the limited ED data points in real time.}
    \label{Fig: Spectral function}
\end{figure*}

In Fig.~\ref{Fig: FGR pi-mode fixed}, we perform 
 numerical FGR calculations for $\xi_\pi = 0.1, 0.2$. This corresponds respectively to the upper and lower right curves in the phase diagram Fig.~\ref{Fig: Decay Channel} and share the same features as the FGR computation with $\xi_0 = 0.1,0.2$ in Fig.~\ref{Fig: FGR 0-mode fixed}. Numerically, one computes the decay rate by truncating the summation over discrete time up to some discrete time $n^*$ in the FGR expressions  \eqref{Eq: FGR of 0 edge mode} and \eqref{Eq: FGR of pi edge mode}. In Fig.~\ref{Fig: Numerical Error}, we show $\Gamma_0$ and $\Gamma_{0,B}$ for different $n^*$ for $gT=1.8$ and $J_xT=2.65$, corresponding to the  rightmost data in the right panel of Fig.~\ref{Fig: FGR pi-mode fixed}. For $\Gamma_0$, it saturates at $n^* \sim 40$ and starts to fluctuate at $n^* \sim 190$. The fluctuation at late times are from revivals in any finite size system. Therefore, we numerically determine the decay rate by setting $n^*$ to be some value before the revivals occur, with $n^*$ varying for different $gT$ and $J_xT$. When the decay rate is small compared to the fluctuation, e.g., for $\Gamma_{0,B}$ in Fig.~\ref{Fig: Numerical Error}, we estimate the decay rate by taking the average from $n^*=100$ to 
 some $n^*$ before revivals occur. As shown in the right panel of Fig.~\ref{Fig: Numerical Error}, the scale of the fluctuation is about $10^{-6}$ and sets the accuracy of the numerical computation. Nevertheless, we still present data points for these small decay rates (smaller than $10^{-6}$) to indicate the existence of a much slower decaying channel.

\section{Spectral function of the autocorrelation function}
\label{Sec:C}

Here we present the spectral function $\Tilde{A}_\infty^x(\omega)$ defined as
\begin{align}
    \Tilde{A}_\infty^x(\omega) = \sum_{n=-\infty}^{n=\infty} A_\infty^x(n)e^{-i\omega nT}.
\end{align}
with $A_\infty^x(-n)=A_\infty^x(n)$. We show the numerical results for the spectral function $\Tilde{A}_\infty^x(\omega)$ in Fig.~\ref{Fig: Spectral function} with the same parameters as Fig.~\ref{Fig: Autocorrelation}. The existence of almost strong $0$ and $\pi$-modes are reflected in the peaks at $\omega = 0$ and $\pi/T$ of the spectral function. The width of these peaks are in principle directly related to the decay rate. Numerically, it is impractical to compute the autocorrelation at every time step $n$. The spectral function is numerically computed from limited ED data points in Fig.~\ref{Fig: Autocorrelation}, which leads to the oscillations in the numerical spectral function. Therefore, it is more convenient to determine the decay rate from rescaling the autocorrelation in the time domain, where only few ED data points can capture the full decay behavior.

\end{document}